\documentclass[reprint,twocolumn,showpacs,showkeys]{revtex4-1}

\usepackage{latexsym,graphicx,graphics}
\usepackage{amsmath,epsfig}

\newcommand{\abs}[1]{\left| #1 \right|}

\newcommand{\bb}[1]{ \mbox{\boldmath$ #1$}}

\newcommand{\bbb}[1]{ \mbox{\boldmath$\bar{ #1}$}}

\newcommand{\Eq}[1]{Eq.~(\ref{#1})}
\newcommand{\Eqs}[2]{Eqs.~(\ref{#1})--(\ref{#2})}

\newcommand{\figstyle}[1]{\small{#1}}

\newcommand{\unit}[1]{\bb{\hat{#1}}}

\begin{document}

\title{Magnetized spiral chains of plasmonic ellipsoids for one-way optical waveguides}
\author{Y. Hadad}
\surname{Hadad}
\email{hadady@eng.tau.ac.il}
\author{Ben Z. Steinberg}
\surname{Steinberg}
\email{steinber@eng.tau.ac.il}
\affiliation{School of Electrical Engineering, Tel Aviv University, Ramat-Aviv, Tel-Aviv 69978  Israel}
\date{July 2010}

 \begin{abstract}
 When a linear chain of plasmonic nano-particles is subject to longitudinal magnetic field, it exhibits optical Faraday rotation. If the magnetized nano-particles are plasmonic ellipsoids arranged as a spiral chain, the interplay between the Faraday rotation and the geometrical spiral rotation (structural chirality) can strongly enhance non-reciprocity. This interplay forms a waveguide that permits one way propagation only, within four disjoint frequency bands; two bands for each direction.
 \end{abstract}


\maketitle

Linear chains of identical and equally spaced micro-particles have been studied in a number of publications \cite{Quinten}-\cite{EnghetaChain}. It has been shown that they allow the propagation of optical modes with relatively low attenuation and with no radiation to the free space. This property is obtained if the inter-particle distance is smaller then the free space wavelength $\lambda$, and then the total width of the modes can be much smaller than $\lambda$. Hence the name "Sub-Diffraction Chains" (SDC). SDCs are potential candidates for dense integration of optical systems, and were proposed as guiding structures, junctions, and couplers \cite{Quinten}-\cite{Lomakin}. Very recently, a spiral SDC was suggested as a chiral waveguide \cite{Lomakin2}.

One way waveguides are another type of key components in optical systems. Beside their potential role in reducing disorder effects and unwanted couplings \cite{YUFAN}, they are used in practice as optical isolators and circulators.

\begin{figure}[htbp]
\vspace*{-0.15in}
    \hspace*{-0.2in}
        \includegraphics[width=8.5cm]{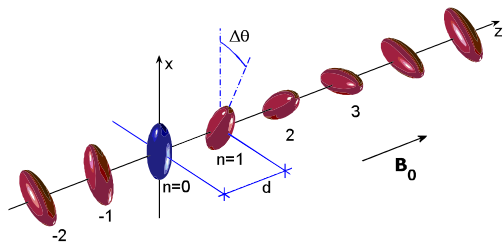}\vspace*{-0.3in}
    \caption{\figstyle{A spiral chain of prolate ellipsoids, subject to a longitudinal magnetic field.}}
    \label{fig1}
\end{figure}

We suggest the SDCs as candidates for one-way waveguides. The propagation in plasmonic particles SDCs subject to a longitudinal magnetic field, is non-reciprocal as it possesses Faraday rotation \cite{Damon1,Damon2}. We suggest an SDC in which the plasmonic particles are \emph{prolate ellipsoids} arranged in a \emph{spiral} structure as in \cite{Lomakin2}, but under longitudinal magnetic field. See Fig.~\ref{fig1}. We show that the interplay between the Faraday rotation and the geometrical spiral rotation (structural chirality) can \emph{strongly enhance non-reciprocity}, and practically forms a one-way waveguide. This one-way waveguide is more compact than the one suggested in \cite{YUFAN} (but perhaps more difficult to fabricate), and requires an order of magnitude weaker magnetic field. At first glance it has a clear physical interpretation; a needle-like particle may act locally as a polarizer in the sense that it is strongly excited by fields aligned with its longest axis, and weakly excited by fields normal to this axis. Therefore the spiral chain of ellipsoids acts as a distributed polarizer that rotates together with the Faraday-rotated SDC mode, doing in a distributed fashion the same job that a couple of polarizers do in a conventional polarization-dependent isolator. However, this physical interpretation is somewhat short of capturing the rich physics conveyed by the two-type rotations interplay. The one-way property exists even with ``footballs'' with axes ratio of 1:2.  Furthermore, a single spiral chain can support one way behavior in \emph{four} disjoint frequency bands simultaneously, two for each direction. The propagation in the ``forbidden'' direction is practically scattered from the chain to the free space; it becomes a radiation mode. Principally these radiation modes can be blocked by a periodic structure as in \cite{YUFAN}.
It should be emphasized that the effects reported here are not restricted to prolate-ellipsoid particles. Thin wire particles or any other particles with geometry that breaks spherical symmetry would convey essentially the same physics; the heart of the matter lies in the \emph{interplay between chirality and Faraday rotation}, and not in the specific form of the ellipsoids. The choice of ellipsoids is made to simplify the mathematical analysis. Structures as in Fig.~\ref{fig1} were fabricated
 using force mediating polymer to bend silicon nanopillars \cite{Scherer}, that can be coated by metal. Bending angle can reach $50-60^\circ$ and it can be tuned by varying, e.g., electron beam exposure. A linear array of vertically grown silicon nanopillars was deformed by bending the pillars sideways, with bending angle that gradually increases along the array, thus producing a chain very similar to Fig.~\ref{fig1} (see Figs.~2,6 in \cite{Scherer}).

To study the system, we use the Discrete Dipole Approximation (DDA) and polarizability theory. These are standard tools used in many works on SDCs \cite{Brongersma}-\cite{Lomakin2}. They hold when the particle diameter $D$ is much smaller than the wavelength, and the inter-particle distance $d$ is large compare to $D$. Studies show excellent agreement with exact solutions even when $d=1.5D$ \cite{MaierKikAtwater}. Finally note that the chain is periodic only for rational $\Delta\theta/2\pi$, but in a reference frame that rotates together with the spiral it is periodic for any $\Delta\theta$. Our analysis does not assume a` priori any periodicity.

  If a small particle with electric polarizability $\bb{\alpha}$ is subject to an exciting electric field  whose local value in the \emph{absence of the particle} is $\bb{E}^L$, its response is described by the electric dipole $\bb{p}=\bb{\alpha}\bb{E}^L$.
The tensor-polarizability of a general ellipsoid made of an anisotropic material $\bb{\epsilon}$ can be found in \cite{SihvolaBook} for the static case. In the dynamic case it needs to be augmented to incorporate radiation loss \cite{RadiativeCorrection}. If the ellipsoid principal axes are aligned with the $x,y,z$ axes, its dynamic polarizability $\bb{\alpha}$ is obtained via
\begin{equation}
\bb{\alpha}^{-1}=\frac{k^3}{6\pi\epsilon_0}\,\left(
\bbb{\alpha}_h^{-1}-i{\bf I}_3\right)\label{eq1}
\end{equation}
where
\begin{equation}
\bbb{\alpha}_h^{-1}=\frac{6\pi}{k^3V}\,
\left[{\bf I}_3+{\bf L}\left({\bb{\epsilon}}-{\bf I}_3\right)\right]\, \left({\bb{\epsilon}}-{\bf I}_3\right)^{-1}.
\label{eq1a}
\end{equation}
 Here $k$ is the free space wavenumber. ${\bf I}_3$ is the $3\times 3$ identity matrix; together with the factor $-i\frac{6\pi\epsilon_0}{k^3}$ it represents the particle radiation loss \cite{RadiativeCorrection}. $\bbb{\alpha}_h$ is the normalized Hermitian matrix representing the ellipsoid geometry and material. $V=4\pi a_xa_ya_z/3$ is the ellipsoid volume and
 $a_x, a_y, a_z$ are its semiaxes. ${\bf L} =\mbox{diag}(N_x,N_y,N_z)$ is the depolarization matrix whose entries are obtained by elliptic integrals and satisfy $\sum_uN_u=1$ \cite{SihvolaBook} . To get a feeling of the numbers involved, note that a sphere of radius $r$ has $a_u=r,\, N_u=1/3$. A \emph{prolate} ellipsoid ($a_x>a_y=a_z=a$), has $N_y=N_z=(1-N_x)/2$, $N_x=(1-e^2)[\ln(\frac{1+e}{1-e})-2e]/(2e^3)$, where $e=\sqrt{1-a^2/a_x^2}$. A ``football'' with $a_x=2a$ has $N_x\approx 0.1736, N_y\approx 0.4132$. A needle with $a_x=10a$ has $N_x\approx 0.02, N_y\approx 0.49$.

If the plasmonic particle is subject to an external DC magnetic field $\bb{B}_0=\unit{z}B_0$, $\bb{\epsilon}$ is that of a magnetized plasma \cite{JACKSON}. Then
\begin{equation}
\bbb{\alpha}_h^{-1}=\left(
\begin{array}{ccc}
\gamma_{xx} & i\gamma_{xy} & 0\\
-i\gamma_{xy} & \gamma_{yy} & 0\\
0 & 0& \gamma_{zz}\end{array}\right)\label{eq2}
\end{equation}
with
\begin{equation}
\begin{array}{ll}
\gamma_{uu}=& \frac{6\pi}{k^3V}\, \left(N_u-\frac{\omega^2}{\omega_p^2}\right)\\
\gamma_{xy}=& \frac{6\pi}{k^3V}\, \frac{\omega\omega_b}{\omega_p^2} \end{array}
\label{eq2a}
\end{equation}
where $\omega_p$ and $\omega_b=-q_eB_0/m_e$ are the plasma and cyclotron frequencies.
Equations (\ref{eq1})-(\ref{eq2a}) describe the ellipsoid at the origin. The $n$-th particle polarizability, $\bb{\alpha}_n$, is
\begin{equation}
\bb{\alpha}_n={\bf T}_{-n}\bb{\alpha}{\bf T}_n, \label{eq3}
\end{equation}
where ${\bf T}_n$ is a rotation matrix, whose non-zero entries are $t_{11}=t_{22}=\cos n\Delta\theta$, $t_{33}=1$, $t_{12}=-t_{21}=\sin n\Delta\theta$.
The electric field at $(0,0,z)$ due to a short dipole $\bb{p}$ at $(0,0,z')$ is given by the matrix relation
\begin{eqnarray}
\bb{E}(z)&=&\epsilon_0^{-1}{\bf A}(z-z') \,\bb{p}\label{eq4a}\\
{\bf A}(z)&=&\frac{e^{ik\abs{z}}}{4\pi\abs{z}}\,\left[k^2{\bf A}_1 + \left(\frac{1}{z^2}-\frac{ik}{\abs{z}}\right){\bf A}_2\right]
\label{eq4b}
\end{eqnarray}
here ${\bf A}_1=\mbox{diag}(1,1,0),\,{\bf A}_2=\mbox{diag}(-1,-1,2)$. We express now the local exciting field of the $m$-th particle in the chain as a sum of contributions from all its neighbor-dipoles, and apply $\bb{\alpha}_m$. The result relates the $m$-th dipole excitation $\bb{p}_m$ to its neighbors
\begin{equation}
\bb{p}_m=\epsilon_0^{-1}{\bf T}_{-m}\bb{\alpha}{\bf T}_m\!\! \sum_{n,\,n\ne m}\!\! {\bf A}[(m-n)d]\, \bb{p}_n. \label{eq6}
\end{equation}
Note that ${\bf T}_{-m}={\bf T}^{-1}_{m}$, ${\bf T}_{m}={\bf T}_{m-n}{\bf T}_{n}$. Also,  ${\bf T}_n$ and ${\bf A}$ commute: ${\bf T}_n{\bf A}(z)={\bf A}(z){\bf T}_n$. Hence we obtain the shift-invariant difference matrix-equation
\begin{equation}
\bb{p}^r_m=\epsilon_0^{-1}\bb{\alpha}\sum_{n,\,n\ne m}{\bf A}[(m-n)d]{\bf T}_{m-n}\,\, \bb{p}^r_n \label{eq7}
\end{equation}
where $\bb{p}^r_n$ is the rotated dipole
\begin{equation}
\bb{p}^r_n\, =\, {\bf T}_n\bb{p}_n.
\label{eq8}
\end{equation}
For infinite chains, every solution of (\ref{eq7}) can always be expressed as
\begin{equation}
\bb{p}^r_n=\bb{p}^r_0e^{i\beta nd}.\label{eq9}
\end{equation}
Substituting it back into \Eq{eq7}, using \Eqs{eq1}{eq2} and rearranging, we get
\begin{equation}
\left(\bbb{\alpha}_h^{-1}-i{\bf I}_3 -{\bf C}\right)\bb{p}^r_0=0,\label{eq10}
\end{equation}
where ${\bf C}$ is a matrix defined by the sum
\begin{equation}
{\bf C}=
\frac{6\pi}{k^3}\sum_{n\ne 0}{\bf A}(nd){\bf T}_ne^{-i\beta nd}.\label{eq10a}
\end{equation}
A non-trivial solution for the vector $\bb{p}^r_0$ exists iff the determinant vanishes. In addition, the structure supports an optical mode that does not radiate to free space iff the radiative-loss part represented by $-i{\bf I}_3$ cancels out, rendering $\beta$ real \cite{EnghetaChain}. This can be achieved only with the help of ${\bf C}$ since $\bbb{\alpha}_h^{-1}$ has a real diagonal. Hence, the guided mode satisfies simultaneously
\begin{eqnarray}
\mbox{Det}\left(\bbb{\alpha}_h^{-1}-i{\bf I}_3 -{\bf C}\right)&=&0,\label{eq11}\\
-\mbox{Im}(\mbox{diag}({\bf C}))&=&{\bf I}_3.
\label{eq11a}
\end{eqnarray}
These equations should be solved for the dispersion curve $\beta(\omega)$. Where, in the parameter space $d,\Delta\theta,\beta,\omega$ should we look for the solution? Equation (\ref{eq7}) observes the system within a reference frame that rotates together with the spiral. In this frame the structure is periodic, and from (\ref{eq9}) $\beta$ is $2\pi/d$ periodic. When observed in the ``lab'' frame, $\beta$ splits to \emph{two} wavenumbers $\beta_\ell=\beta\pm\Delta\theta/d$. This is easily verified: apply ${\bf T}_{-n}$ to \Eq{eq9}, and use ${\bf T}_{-n}$ eigenvalues and eigenvectors $\lambda_{1,2}=\exp(\pm in\Delta\theta)$ and $v_{1,2}=(1,\pm i,0)/\sqrt{2}$ ($\lambda_3,v_3$ are ignored as we look for transverse excitations). The result is
\begin{equation}
\bb{p}_n=(v_1^*\cdot\bb{p}_0^r)v_1e^{in(\beta d+\Delta\theta)}+(v_2^*\cdot\bb{p}_0^r)v_2e^{in(\beta d-\Delta\theta)}.\label{eq12}
\end{equation}
In the lab frame, any SDC guided mode must have a real wavenumber within the interval $k<\abs{\beta_\ell}<\pi/d$ \cite{EnghetaChain}, establishing our search domain $kd<\abs{\beta d\pm\Delta\theta}<\pi$. The entries of ${\bf C}$ can be evaluated efficiently using the polylogarithm functions as in \cite{EnghetaChain}, with appropriate $\pm\Delta\theta$ shifts (we skip the details as they pertain only to numerical efficiency). Finally, note that $\bb{\alpha}$ of a prolate ellipsoid has two resonant frequencies, due to the two different axes. Hence, an ellipsoids SDC possesses \emph{two} frequency bands of guided modes. As we show below each of the bands supports \emph{two} one-way waveguides.

\begin{figure}[htbp]
    \centering
        \includegraphics[width=7.5cm]{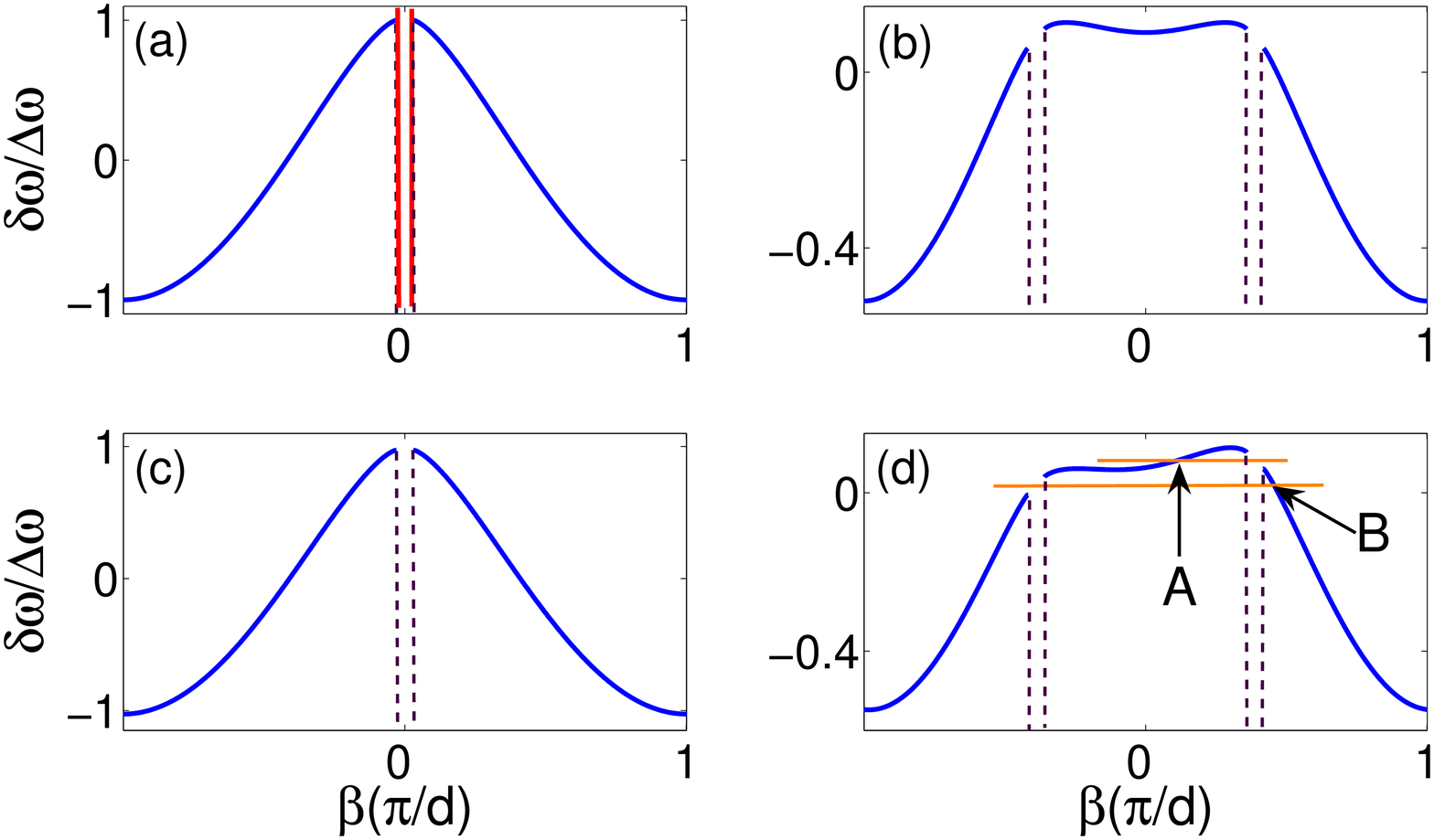}
    \caption{\figstyle{The chain dispersion in the lower mode. Blue lines are guided modes and black-dashed lines are radiation modes. (a) all ellipsoids are lined up ($\Delta\theta=0$) and $\bb{H}_0=0$. Note the gap at $\abs{\beta}<k$, confined between the two radiation modes. Light lines are shown in red. (b) $\Delta\theta=70^\circ$ and $\bb{H}_0=0$. The gap splits symmetrically. (c) The same as (a) but with a magnetic field of $\omega_b=0.01\omega_p$. (d) A spiral with $\Delta\theta=70^\circ$, and magnetic field of $\omega_b=0.01\omega_p$. The gaps of (b) shift to non-symmetrical locations. One-way guiding in $+z$ ($-z$) direction is obtained in point A (B).}}
    \label{fig2}
\end{figure}

We solved (\ref{eq11})-(\ref{eq11a}) for $\omega(\beta)$, with the following parameters. The ellipsoid axes are $a_x=0.25d, a_y=a_z=0.5a_x$ and $d=\lambda_p/30$, where $\lambda_p$ is the $\omega_p$ wavelength. The results are shown in Fig.~\ref{fig2}. 
When no magnetic field is present and all ellipsoids are lined up ($\Delta\theta=0$), they form an SDC with transmission bandwidth $\Delta\omega_1=3.229\times 10^{-3}\omega_p$ around the central frequency $\omega_{0\, 1}=0.41702\omega_p$ - the lower band, and a second SDC with $\Delta\omega_2=2.05\times 10^{-3}\omega_p$ around the central frequency $\omega_{0\, 2}=0.643052\omega_p$ - the upper band. Both have dispersion properties similar to the transverse mode in \cite{EnghetaChain}. For these parameters, the DDA works very well \cite{MaierKikAtwater}. The lower band is shown in Fig.~\ref{fig2}a. The vertical axis is $\delta\omega=\omega-\omega_{0\, 1}$, normalized to $\Delta\omega_1$. Note the gap at $\abs{\beta}d<k$. The dispersion lines shown as dashed lines at the edges of this gap are practically touching the light-line cone $\abs{\beta}=k$ (shown in red), representing plane-wave modes that practically do not interact with the chain \cite{EnghetaChain}, and can be considered as radiation modes. In Fig.~\ref{fig2}b a spiral rotation of $\Delta\theta=70^\circ$ is added to the chain, yet with $\bb{B}_0=0$. The gap splits in two symmetrically. For clarity, the nearly vertical lines of the radiation modes were omitted. In Fig.~\ref{fig2}c the ellipsoids are lined up ($\Delta\theta=0$), with a magnetic field of $\omega_b=0.01\omega_p$ . The dispersion curve is symmetric, but a solution for $\bb{p}_0$ reveals Faraday rotation. Finally, Fig.~\ref{fig2}d shows the dispersion for $\Delta\theta=70^\circ$ and $\omega_b=0.01\omega_p$. The magnetic field shifts the two gaps of (b), formed by the spiral rotation, to non-symmetrical locations. At point A ($\omega=\omega_A$, $\delta\omega=0.0866\Delta\omega$) the chain permits propagation only in the $+z$ direction (positive group velocity). Formally, the horizontal line of point A intersects also the light-cone that represents propagation in the opposite direction. However this line represents a background radiation mode that is practically not excited as its interaction with the chain is very weak. Likewise, at point B ($\omega=\omega_B$, $\delta\omega=0.0278\Delta\omega$) the waveguide permits propagation only in the $-z$ direction. Two one-way guiding are supported, one for each direction.

\begin{figure}[htbp]
    \centering
        \includegraphics[width=7.5cm]{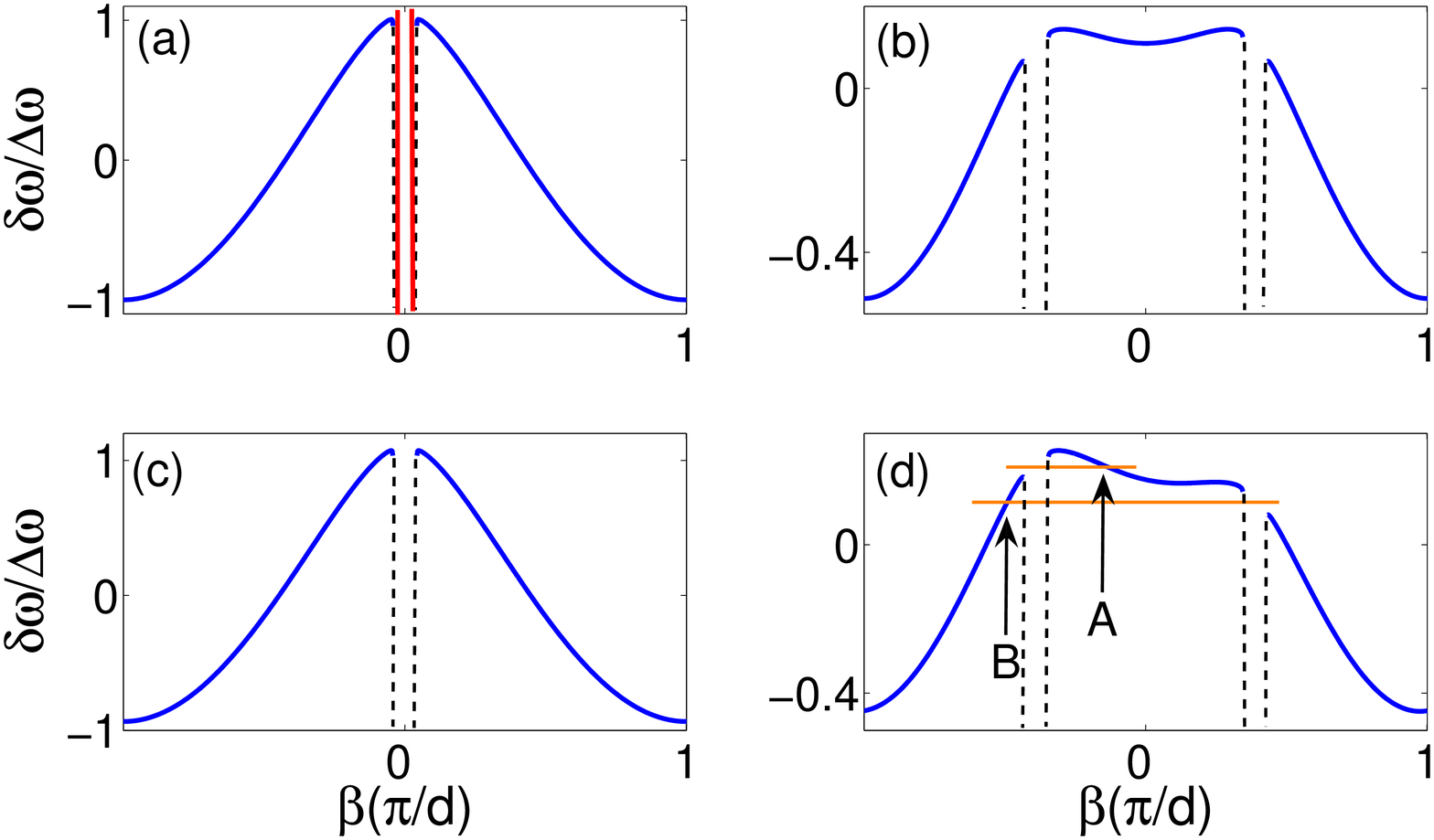}
    \caption{\figstyle{The same as Fig.~\ref{fig2} but for the higher mode. Note the anti-symmetry with respect to the lower mode: one-way $-z$ ($+z$) is obtained in point A (B).}}
    \label{fig3}
\end{figure}

Essentially the same holds also for the SDC modes in the upper band, where we have $\delta\omega=0.1944\Delta\omega$ for point A, allowing only $-z$ propagation, and $\delta\omega=0.111\Delta\omega$ at point B, allowing only $+z$ propagation. This is shown in Fig.~\ref{fig3}.

\begin{figure}[htbp]
    \centering
        \includegraphics[width=7.5cm]{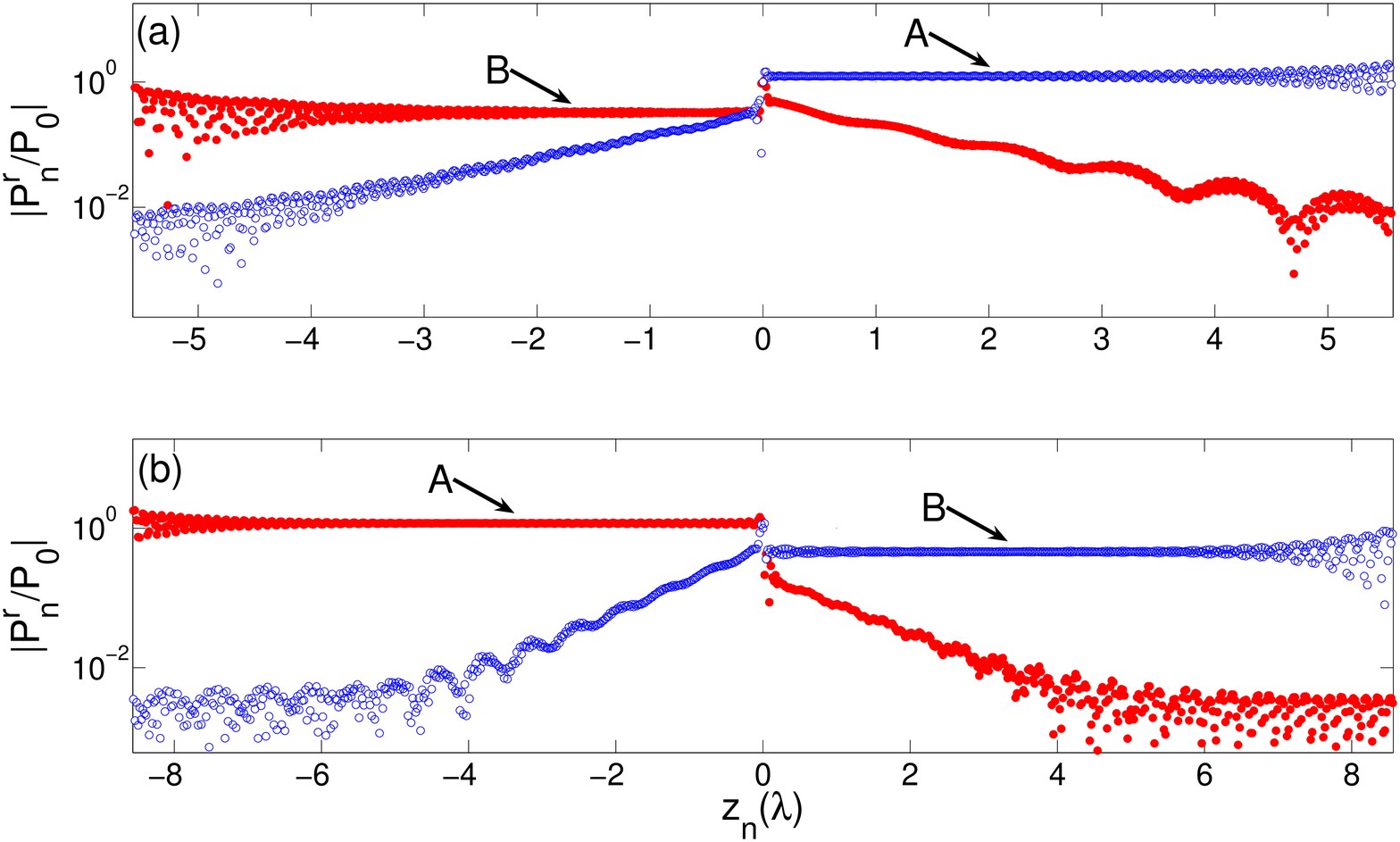}\vspace*{-0.2in}
    \caption{\figstyle{The chain response to a unit excitation of the ellipsoid at the origin. (a) Lower mode. Curves A, B correspond to $\omega=\omega_A,\omega_B$, respectively. (b) Upper band.}}
    \label{fig4}
\end{figure}

To verify these properties, we have simulated a finite chain of $N=800$ ellipsoids with the above parameters. The ellipsoid at the origin is forced by a unit amplitude. When hundreds of deep sub-wavelength particles are considered, full 3D solutions based e.g. on FDTD technique would require tens of thousands of grid-points and may become too heavy to handle; an equivalent model that encapsulates the main physics, with considerably fewer unknowns is called for.
 Recall that for the parameters used here the DDA is usually in excellent agreement with exact full-wave 3D numerical solutions \cite{MaierKikAtwater}, whenever the latter can be practically applied. Hence we use the DDA also here. Then
\Eq{eq7} still applies, but with two minor changes. The summation becomes finite so the shift-invariance is lost. Also, the excitation of the forced particle at the origin is known: $\bb{p}_0^r=\unit{x}$. Thus \Eq{eq7} becomes a finite matrix equation for $N-1$ unknowns $\bb{p}_n^r,\, n\ne 0$, that is solved numerically. Figure \ref{fig4}a shows $\abs{\bb{p}_n^r}$ on a logarithmic scale for excitation frequencies $\omega_A$ $\omega_B$ at the lower band, and Fig.~\ref{fig4}b shows the same but at the higher band.
In all cases, propagation in the ``forbidden'' direction decays within a distance of $O(\lambda)$, becoming at least two orders of magnitude weaker. Since the simulated chain is of finite length, a reflected mode is excited whenever the propagating mode hits the chain end. This mode interferes with the propagating mode, and generates strong oscillations.  However, since the reflected mode is propagating in the ``forbidden'' direction, it decays exponentially, and so do the oscillations. This is seen clearly in the figures.

\end{document}